\documentclass[12pt]{article}

\usepackage{amsmath}
\usepackage{xcolor}
\usepackage{subfigure}
\usepackage{appendix,epic,eepic,amsmath,amsthm,amssymb,epsfig,cite,indentfirst,oldgerm}
\renewcommand{\theequation}{\arabic{equation}}
\newcommand{\EQ}{\begin{equation}}
\newcommand{\EN}{\end{equation}}

\newcommand{\bear}{\begin{eqnarray}}
\newcommand{\ear}{\end{eqnarray}}
\newcommand{\bt} { \begin{tabular} }
\newcommand{\et}{ \end{tabular} }
\newcommand{\bc} { \begin{center} }
\newcommand{\ec}{ \end{center} }

\newcommand{\btb} { \begin{table} }
\newcommand{\etb}{ \end{table} }

\begin{document}

\topmargin 0pt
\oddsidemargin 5mm
\newcommand{\NP}[1]{Nucl.\ Phys.\ {\bf #1}}
\newcommand{\PL}[1]{Phys.\ Lett.\ {\bf #1}}
\newcommand{\NC}[1]{Nuovo Cimento {\bf #1}}
\newcommand{\CMP}[1]{Comm.\ Math.\ Phys.\ {\bf #1}}
\newcommand{\PR}[1]{Phys.\ Rev.\ {\bf #1}}
\newcommand{\PRL}[1]{Phys.\ Rev.\ Lett.\ {\bf #1}}
\newcommand{\MPL}[1]{Mod.\ Phys.\ Lett.\ {\bf #1}}
\newcommand{\JETP}[1]{Sov.\ Phys.\ JETP {\bf #1}}
\newcommand{\TMP}[1]{Teor.\ Mat.\ Fiz.\ {\bf #1}}

\renewcommand{\thefootnote}{\fnsymbol{footnote}}

\newpage
\setcounter{page}{0}
\begin{titlepage}
\begin{flushright}

\end{flushright}
\vspace{0.25cm}
\begin{center}
{\large The spectrum properties of an integrable $G_2$ invariant vertex model.} \\
\vspace{1cm}
{\large M.J. Martins } \\
\vspace{0.1cm}
{\em Universidade Federal de S\~ao Carlos\\
Departamento de F\'{\i}sica \\
C.P. 676, 13565-905, S\~ao Carlos (SP), Brazil}\\
\vspace{0.2cm}
\end{center}
\vspace{0.1cm}

\begin{abstract}
This paper is concerned with the study of properties of the exact solution of the fundamental
integrable $G_2$ vertex model. The model $R$-matrix and respective spin chain 
are presented in terms of the basis generators of the $G_2$ Lie algebra. This formulation permits us
to related the number of the Bethe roots of the respective Bethe equations with the eigenvalues
of the $U(1)$ conserved charges from the Cartan subalgebra of $G_2$. The Bethe equations
are solved by a peculiar string structure which combines complex three-strings with real roots 
allowing us determine the bulk properties in the thermodynamic limit. We argue that $G_2$ spin chain is gapless but
the low-lying excitations have two different speeds of sound and the underlying continuum limit is therefore not 
strictly Lorentz invariant.
We have investigate
the finite-size corrections to the
ground state energy and proposed that the critical properties of the 
system should be governed
by the product of two $c=1$ conformal field theories. By combining numerical and analytical methods we  have computed  
the bulk free-energy of the $G_2$ vertex model.
We found that there are three regimes in the spectral parameter in which
the free-energy is limited and continuous. There exists however at least two sharp corner
points in which the bulk-free energy is not differentiable.
\end{abstract}

\vspace{.15cm} \centerline{}
\vspace{.1cm} \centerline{Keywords: Integrability, $G_2$ vertex model, $G_2$ spin chain .}
\vspace{.15cm} \centerline{February 2023}

\end{titlepage}


\pagestyle{empty}

\newpage

\pagestyle{plain}
\pagenumbering{arabic}

\renewcommand{\thefootnote}{\arabic{footnote}}
\newtheorem{proposition}{Proposition}
\newtheorem{pr}{Proposition}
\newtheorem{remark}{Remark}
\newtheorem{re}{Remark}
\newtheorem{theorem}{Theorem}
\newtheorem{theo}{Theorem}

\def\ll{\left\lgroup}
\def\rr{\right\rgroup}

\newtheorem{Theorem}{Theorem}[section]
\newtheorem{Corollary}[Theorem]{Corollary}
\newtheorem{Proposition}[Theorem]{Proposition}
\newtheorem{Conjecture}[Theorem]{Conjecture}
\newtheorem{Lemma}[Theorem]{Lemma}
\newtheorem{Example}[Theorem]{Example}
\newtheorem{Note}[Theorem]{Note}
\newtheorem{Definition}[Theorem]{Definition}

\section{Introduction}

Integrable two-dimensional vertex models are important paradigms of exactly solved models
of statistical mechanics \cite{BAX}. Their statistical weights are directly related to 
a central object in the theory 
of quantum integrable systems named the $R$-matrix. For the simplest models
the $R$-matrix acts on the product of two equidimensional spaces $V \otimes V$  
depending on a single spectral parameter $\lambda$. The integrability condition 
is assured
by imposing the Yang-Baxter equation,
\begin{equation}
\label{YB}
\mathrm{R}_{12}(\lambda_1-\lambda_2)
\mathrm{R}_{13}(\lambda_1)
\mathrm{R}_{23}(\lambda_2)=
\mathrm{R}_{23}(\lambda_2)
\mathrm{R}_{13}(\lambda_1)
\mathrm{R}_{12}(\lambda_1-\lambda_2),
\end{equation}
where $\mathrm{R}_{ab}(\lambda) \in V_a \otimes V_b$. 

An important family of $R$-matrices are those based on the fundamental 
representation of each simple Lie algebra ${\cal G}$, see for instance \cite{YAN,GAU,ZAMO,KUL1,OGI}.
The $R$-matrix elements 
are expressed in terms
of polynomials on the spectral parameter $\lambda$ and they are often called rational $R$-matrices. 
For a given rational $R$-matrix one can built the generator of the conserved 
quantities associated 
to the lattice vertex model.
This operator is called the 
transfer matrix and will be denoted by $\mathrm{T}_{L}(\lambda)$ where $L$ is the system size. The transfer matrix 
can be represented 
as trace of a ordered product 
of $R$-matrices \cite{BAX,FAD},
\begin{equation}
\label{TRA}
\mathrm{T_{L}}(\lambda)=\mathrm{Tr}_0\left[ 
\mathrm{R}_{01}(\lambda)
\mathrm{R}_{02}(\lambda) \dots 
\mathrm{R}_{0\mathrm{L}}(\lambda)\right], 
\end{equation}
where 
the trace is taken on the $0$-th auxiliary space.

In order to extract the physical properties of the system 
in the thermodynamic
limit $L \rightarrow \infty $ it is necessary the knowledge of 
the structure of the eigenvalues
of the transfer matrix $\mathrm{T}_{L}(\lambda)$. This step has been successfully 
achieved for many of fundamental
${\cal G}$ invariant vertex models either 
by means of phenomenological method denominated 
analytical Bethe ansatz \cite{OGIV,RES1,RES2,JUN} or 
through the framework of the quantum inverse  scattering \cite{KOR,DEV,KUL,MARA}.
However, as far as we know, these solutions have been so far explored to determine 
the thermodynamic limit properties 
of the vertex models based mainly on classical Lie algebras, see for instance \cite{VEGA,VESU,IZER,DELO}.
It appears that the behaviour of the free-energy of
the majority of vertex models based on the exceptional Lie algebras
as well as the nature of the excitations of the
underlying spin chains have not yet been investigated in the literature.
In this paper we hope to start to bridge this gap by 
investigating the thermodynamic limit properties  of the 
$G_2$ invariant vertex model. Recall here that this symmetry is the smallest possible 
exceptional Lie algebra
besides the automorphism group of the algebra of octonions. The dimension of the Hilbert space
permits us to study the spectrum numerically for moderate lattice sizes making it possible
the identification of the structure of Bethe roots associated to the ground state. This step
is essential to determine the infinite volume properties of the vertex model by the means
of Bethe ansatz techniques.

This paper have been organized as follows.  We start next section 
by discussing basic properties of the
seven dimensional representation of $G_2$ Lie algebra. This allows us to represent 
the $R$-matrix 
and the corresponding spin chain Hamiltonian  
in terms of combinations of the $G_2$ quadratic Casimir operator. We revisit the exact solution
of the $G_2$ vertex model by a coupled set of Bethe ansatz equations. In particular, we relate the number of
the Bethe roots in each level with 
the eigenvalues 
of the $U(1)$ conserved charges from the Cartan subalgebra of $G_2$. To the best of our knowledge the
above presentation of the $G_2$ vertex model properties is new in the literature.
In section 3 we investigate the
ground state properties of the $G_2$ spin chain and found that the respective Bethe roots 
combine complex three-strings with real roots. This permits us to study the thermodynamic limit
and the spin chain ground state energy per site is computed exactly. In section 4 we argue that the low-lying 
excitations over 
the ground state are
gapless and that they travel with two different sound velocities. 
The continuum limit of the system is therefore expected to be described in terms
of the product of two conformal field theories. We next have investigated the finite-size 
corrections 
to the ground state and to the lowest excitation of the $G_2$ spin chain. 
Our results are compatible with the interpretation that the underlying 
conformal field theories should have 
both central charge $c=1$. The lowest excitation carries non zero momenta and our finite-size analysis
indicates that the conformal dimension associated to this excitation 
should be $X=1$.
In section 5 we compute the bulk free-energy of the $G_2$ vertex model in a  
region of the spectral parameter in which this function is limited. We
find that the
bulk free-energy is a continuous function on the spectral parameter 
but there exists  at least two sharp corner 
points in which the free-energy is not differentiable. Section 6 is reserved for our
conclusions and in Appendix A we present technical details omitted in the text.

\section{The $G_2$ vertex model}

In this section we shall present the $R$-matrix of the vertex model in terms 
of the generators
of the $G_2$ symmetry.  To this end we start by discussing some basic properties  
of the $G_2$ Lie algebra. This algebra is 
a fourteen dimensional algebra of rank two \cite{HUM,JAR} and here we denote  
the respective 
Cartan-Weyl basis by,
\begin{equation}
\{ H_1, H_2, E_1, \cdots, E_6, E_1^{\dagger}, \cdots, E_6^{\dagger} \}
\end{equation}
where the subset $\{H_1,H_2\}$ forms the Cartan subalgebra of $G_2$. 

We now consider the task of presenting explicit expressions for  such
generators in the case of the seven
dimensional representation. 
We first represent
the generators of the two-dimensional 
Cartan subalgebra 
as a linear combination
of two independent diagonal operators,
\begin{equation}
H_1= \widehat{H}_1+ \widehat{H}_2,~~~H_2=\frac{1}{\sqrt{3}}(\widehat{H}_1-\widehat{H}_2)
\end{equation}
where the auxiliary operators $\widehat{H}_1$ and $\widehat{H}_2$ are,
\begin{equation}
\widehat{H}_1=\mathrm{diag}\{1,0,1,0,-1,0,-1\},~~~
\widehat{H}_2=\mathrm{diag}\{0,1,-1,0,1,-1,0\},~~~
\end{equation}

We next choose the two simple roots associated to the $G_2$ Lie algebra by 
the vectors $\alpha_1=(-1,-\frac{1}{\sqrt{3}})$ and 
$\alpha_2=(2,0)$. The corresponding roots generators are, 
\begin{equation}
E_1=\frac{1}{\sqrt{3}}\left( \begin{array}{ccccccc} 
	0 &  0 & 0 & 0 & 0 & 0 & 0 \\
	0 &  0 & 0 & 0 & 0 & 0 & 0 \\
	0 &  0 & 0 & 0 & 0 & 0 & 0 \\
	2 &  0 & 0 & 0 & 0 & 0 & 0 \\
	0 &  -\sqrt{2} & 0 & 0 & 0 & 0 & 0 \\
	0 &  0 & -\sqrt{2} & 0 & 0 & 0 & 0 \\
	0 &  0 & 0 & 2 & 0 & 0 & 0 \\
	\end{array}
	\right),~~
E_2=\sqrt{2}\left( \begin{array}{ccccccc} 
	0 &  0 & 0 & 0 & 0 & 1 & 0 \\
	0 &  0 & 0 & 0 & 0 & 0 & 1 \\
	0 &  0 & 0 & 0 & 0 & 0 & 0 \\
	0 &  0 & 0 & 0 & 0 & 0 & 0 \\
	0 &  0 & 0 & 0 & 0 & 0 & 0 \\
	0 &  0 & 0 & 0 & 0 & 0 & 0 \\
	0 &  0 & 0 & 0 & 0 & 0 & 0 \\
	\end{array}
	\right)
\end{equation}

The expression of the other four generators can 
be obtained by exploring the   
commutation relations of $G_2$. In particular, we have the following relations, 
\begin{equation}
E_3=\frac{1}{\sqrt{2}}[E_1,E_2],~~~E_4=\sqrt{\frac{3}{8}}[E_1,E_3],~~E_5=\frac{1}{\sqrt{2}}[E_1,E_4],~~E_6=\frac{1}{\sqrt{2}}[E_2,E_5]
\end{equation}

The set of two Cartan generators and six roots generators together 
with their transposes can be used to compute the quadratic 
Casimir operator \cite{HUM,JAR}. The expression 
of this operator
acting on pair of sites $(i,i+1)$ of an
one dimensional lattice is,
\begin{equation}
C_{i,i+1} =  H_1^{(i)} \otimes H_1^{(i+1)} + H_2^{(i)} \otimes H_2^{(i+1)} + \sum_{l=1}^{6} \left( E_{l}^{(i)} \otimes [E_{l}^{(i+1)}]^{\dagger} +
[E_{l}^{(i)}]^{\dagger} \otimes E_{l}^{(i+1)}  \right)
\end{equation}

We remark that degeneracies of the eigenvalues of the quadratic Casimir operator are compatible with the 
Clebsch-Gordon decomposition $7 \otimes 7= 1 \oplus 7 \oplus 14 \oplus 27$. It is well known that these four moduli correspond 
respectively to the identity, fundamental,
adjoint and non-fundamental representations, see e.g \cite{OKU}.  It turns out that the $R$-matrix of the $G_2$ vertex 
model can be expressed 
in terms of the projective operators onto such irreducible subspaces \cite{OGI}, 
\begin{equation}
\label{RMA}
R_{12}(\lambda)=\rho_0(\lambda)\left(P_{12}^{(27)}+\frac{\lambda-1}{\lambda+1}P_{12}^{(14)} +\frac{\lambda-4}{\lambda+4} P_{12}^{(7)} + \frac{(\lambda-1)(\lambda-6)}{(\lambda+1)(\lambda+6)} P_{12}^{(0)} \right)
\end{equation}
where $\rho_0(\lambda)$ is an arbitrary normalization. Here we fix the normalization by
canceling the poles of the $R$-matrix,
\begin{equation}
\rho_0(\lambda)=(1+\lambda)(4+\lambda)(6+\lambda)
\end{equation}

The underlying $G_2$ symmetry of the $R$-matrix becomes explicitly manifested by writing 
the corresponding projectors in terms of the 
Casimir operator. Their expressions are given by,
\begin{eqnarray}
P_{12}^{(0)} &=& \frac{1}{56} C_{12} -\frac{1}{112} C_{12}^2-\frac{3}{896}C_{12}^3, \nonumber \\
P_{12}^{(7)} &=& -\frac{1}{8} C_{12} +\frac{5}{64} C_{12}^2+\frac{3}{256}C_{12}^3, \nonumber \\
P_{12}^{(14)} &=& \mathrm{Id}_1\otimes \mathrm{Id}_2-\frac{3}{8} C_{12} -\frac{1}{4} C_{12}^2-\frac{3}{128}C_{12}^3, \nonumber \\
P_{12}^{(27)} &=& \frac{27}{56} C_{12} +\frac{81}{448} C_{12}^2+\frac{27}{1792}C_{12}^3 
\end{eqnarray}
where the symbol $\mathrm{Id}_{i}$ denotes the action 
of the seven dimensional identity at the $i$-th site.

Finally, we mention that the above $R$-matrix besides the 
Yang-Baxter equation (\ref{YB})
satisfies other properties of two-dimensional 
factorized scattering theory, namely
\begin{eqnarray}
&&\mathrm{Unitarity:}~~~~~~~~~~~~~~~~~~~~R_{12}(\lambda) R_{21}(-\lambda)= \rho_0(\lambda) \rho_0(-\lambda) \mathrm{Id}_{1} \otimes \mathrm{Id}_{2} \nonumber \\
&&\mathrm{Parity~~invariance:}~~~~~~~~P_{12}R_{12}(\lambda) P_{12}= R_{12}(\lambda) \\
&&\mathrm{Temporal~~invariance:}~~~R_{12}(\lambda)^{t_1 t_2}= R_{12}(\lambda) \\
&&\mathrm{Crossing~~symmetry:}~~~~R_{12}(\lambda)=-V_1 R_{12}(-6-\lambda)^{t_2} V_1^{-1}
\end{eqnarray}
where the symbol $t_{\alpha}$ denotes transposition on the $\alpha$-the space, 
$P_{12}$ is the standard permutation operator 
and the crossing matrix $V$ is anti-diagonal 
whose non-null entries are $V_{i,j}=(-1)^{i-1} \delta_{i,8-j}$.

\subsection{The exact solution}

We start by discussing the form of the eigenvalues for the transfer matrix of the vertex model 
invariant by the $G_2$ Lie algebra.
The eigenvalues was first conjectured in \cite{OGIV,RES2} exploiting the relationship between 
the structure of the Bethe equations with respective Lie algebra Dynkin diagram. This solution was confirmed
by the analytical Bethe ansatz method \cite{JUN} in the context of the quantum group $U_q[G_2]$ generalization 
of the vertex model \cite{KUN}. We recall that this problem has also been investigated in the case of open
boundary conditions for the $U_q[G_2]$ vertex model \cite{BAT}.
In what follows we present the form
of such eigenvalues making explicit the underlying crossing symmetry of the $G_2$ vertex model.
It turns out that the eigenvalues $\Lambda(L,\lambda)$ of the transfer matrix (\ref{TRA},\ref{RMA}) 
can be written in the following form,
\begin{equation}
\label{ein1}
\Lambda(L,\lambda)= \sum_{l=1}^{3}\left[a_l(\lambda)\right]^L G_{l}\left(\lambda,\{\lambda_j^{\alpha}\}\right)
+\left[a_4(\lambda)\right]^L G_{4}\left(\lambda,\{\lambda_j^{\alpha}\}\right)
+\sum_{l=1}^{3}\left[-a_l(-\lambda-6)\right]^L G_{l}\left(-6-\lambda,\{-\lambda_j^{\alpha}\}\right)
\end{equation}
where the leading polynomials $a_{l}(\lambda)$ are given by,
\begin{equation}
\label{ein2}
a_1(\lambda)=(\lambda+1)(\lambda+4)(\lambda+6),~~a_2(\lambda)=a_3(\lambda)=\lambda(\lambda+4)(\lambda+6),~~a_4(\lambda)=\lambda(\lambda+3)(\lambda+6)
\end{equation}

The corresponding dressing polynomials 
$G_{l}(\lambda,\{ \lambda_j^{(\alpha)} \})$ are parametrized in terms of two sets of Bethe roots 
$\{ \lambda_j^{(1)} \}$  and
$\{ \lambda_j^{(2)} \}$ and their expressions are,
\begin{equation}
\label{ein3}
G_{l}(\lambda,\{ \lambda_j^{(\alpha)} \})= \begin{cases}
\displaystyle \prod_{j=1}^{N_{1}}
  \frac{\lambda_{j}^{(1)}+i\lambda-\frac{i}{2}}
 {\lambda_{j}^{(1)}+i\lambda+\frac{i}{2}},~~~~~~~~~~~~~~~~~~~~~~~~~~~~~~\mathrm{for}~~l=1, \vspace{0.2cm} \\
\displaystyle \prod_{j=1}^{N_{1}}
  \frac{\lambda_{j}^{(1)}+i\lambda+\frac{3i}{2}}
 {\lambda_{j}^{(1)}+i\lambda+\frac{i}{2}}
\displaystyle \prod_{j=1}^{N_{2}}
  \frac{\lambda_{j}^{(2)}+i\lambda-i}
 {\lambda_{j}^{(2)}+i\lambda+2i},~~~~~~~\mathrm{for}~~l=2,\vspace{0.2cm} \\
\displaystyle \prod_{j=1}^{N_{1}}
  \frac{\lambda_{j}^{(1)}+i\lambda+\frac{3i}{2}}
 {\lambda_{j}^{(1)}+i\lambda+\frac{7i}{2}}
\displaystyle \prod_{j=1}^{N_{2}}
  \frac{\lambda_{j}^{(2)}+i\lambda+5i}
 {\lambda_{j}^{(2)}+i\lambda+2i},~~~~~~~\mathrm{for}~~l=3,\vspace{0.2cm} \\
\displaystyle \prod_{j=1}^{N_{1}}
  \left(\frac{\lambda_{j}^{(1)}+i\lambda+\frac{3i}{2}}
 {\lambda_{j}^{(1)}+i\lambda+\frac{5i}{2}} \right)
  \left( \frac{\lambda_{j}^{(1)}+i\lambda+\frac{9i}{2}}
 {\lambda_{j}^{(1)}+i\lambda+\frac{7i}{2}} \right)~~\mathrm{for}~~l=4,\vspace{0.2cm} \\
\end{cases}
\end{equation}
while the respective roots satisfy the a system 
of nested Bethe ansatz equations given by,
\begin{eqnarray}
&& \left[\frac{\lambda_{j}^{(1)}+\frac{i}{2}}
 {\lambda_{j}^{(1)}-\frac{i}{2}}\right]^{{L}} =
\prod_{\stackrel{k=1}{k \neq j}}^{N_{1}}
  \frac{\lambda_{j}^{(1)}-\lambda_{k}^{(1)}+i}
 {\lambda_{j}^{(1)}-\lambda_{k}^{(1)}-i}
\prod_{k=1}^{N_{2}}
  \frac{\lambda_{j}^{(1)}-\lambda_{k}^{(2)}-\frac{3i}{2}}
 {\lambda_{j}^{(1)}-\lambda_{k}^{(2)}+\frac{3i}{2}},~~j=1,\dots,N_1, \vspace{0.2cm} 
\nonumber \\
&& \prod_{\stackrel{k=1}{k \neq j}}^{N_{2}}
  \frac{\lambda_{j}^{(2)}-\lambda_{k}^{(2)}+3i}
 {\lambda_{j}^{(2)}-\lambda_{k}^{(2)}-3i}= 
\prod_{k=1}^{N_{1}}
  \frac{\lambda_{j}^{(2)}-\lambda_{k}^{(1)}+\frac{3i}{2}}
 {\lambda_{j}^{(2)}-\lambda_{k}^{(1)}-\frac{3i}{2}},~~~~~~~~~~~~~~~~j=1,\dots,N_2 
\label{bethe}
\end{eqnarray}

A special family of states arise when the corresponding Bethe roots are 
invariant under reflection through the origin, i.e. $\{ \lambda_j^{(\alpha)} \} = 
\{-\lambda_j^{(\alpha)} \}$. 
For $L$ even, we observe that 
the eigenvalues 
(\ref{ein1}-\ref{ein3}) of such states are clearly invariant 
under the crossing spectral translation. Therefore, their eigenvalues are expected to
satisfy the constraint,
\begin{equation}
\label{cross}
\Lambda(L,\lambda)= \Lambda(L,-\lambda-6)
\end{equation}

Considering the expansion of
the logarithmic derivative of the transfer matrix around the regular point $\lambda=0$ we obtain the
respective local integrals of motion. The first non-trivial charge produces the Hamiltonian of the
integrable $G_2$ invariant spin chain. Its expression in terms of the Casimir
operator is,
\begin{equation}
\label{HAM}
{\cal{H}}= \sum_{i=1}^{L} \left(
\frac{37}{48} C_{i,i+1} +\frac{185}{384} C_{i,i+1}^2+\frac{25}{512}C_{i,i+1}^3 -\frac{7}{12} \mathrm{Id}_{i} \otimes \mathrm{Id}_{i+1} \right)
\end{equation}
where periodic boundary condition is assumed.  The corresponding eigenenergies are obtained from the Bethe roots by using
the following equation,
\begin{equation}
E(L)= -\sum_{j=1}^{N_1}
\frac{1}{\left(\lambda_j^{(1)}\right)^2+1/4} +\frac{17}{12}L
\label{ener}
\end{equation}

We now complete our discussion connecting the number of the Bethe roots in each level 
with the conserved $U(1)$ charges of the Cartan subalgebra of $G_2$. First we note that 
the Hamiltonian 
(\ref{HAM}) commutes
with the charges generated by the auxiliary operators 
$\widehat{H}_1$ and $\widehat{H}_2$, namely
\begin{equation}
\label{char}
[{\cal{H}}, \sum_{i=1}^{L} \widehat{H}_1^{(i)}]=
[{\cal{H}}, \sum_{i=1}^{L} \widehat{H}_2^{(i)}]=0
\end{equation}

In particular, the underlying $SU(2)$ symmetry present in the $G_2$ spin chain 
can be expressed directly in terms of a linear combination 
of the above two charges. In fact,
the corresponding spin-$3$ azimuthal operator is,
\begin{equation}
S^{z} =
\sum_{i=1}^{L} \left(3\widehat{H}_1^{(i)}+
2\widehat{H}_2^{(i)}\right)
\end{equation}

As a consequence of that the Hilbert space can separated into 
disjoint sectors labeled 
by the eigenvalues of the conserved
charges defined by Eq.(\ref{char}). It is possible to label these eigenvalues in terms 
of two quantum numbers $m_1$ and $m_2$ satisfying the 
following constraints,
\begin{eqnarray}
0 \leq m_1 \leq L,&&~~~ -L \leq m_2 \leq L-m_1 \nonumber \\
-L \leq m_1 \leq 0,&&~~~ -m_1-L \leq m_2 \leq L
\end{eqnarray}

Due to the spin-reversal symmetry the positive and negative 
eigenvalues of the azimuthal spin-$3$ operator are degenerated. By restricting 
to the sectors
with positive eigenvalues we find that the number of Bethe roots can be 
parametrized as,
\begin{equation}
N_1=2L -2m_1-m_2,~~~~N_2=L-m_1-m_2
\end{equation}

We now turn to the analysis of the ground state properties of the $G_2$ 
invariant spin chain (\ref{HAM}).

\section{The spin chain ground state}

\begin{figure}[ht]
\begin{center}    
\includegraphics[width=13cm]{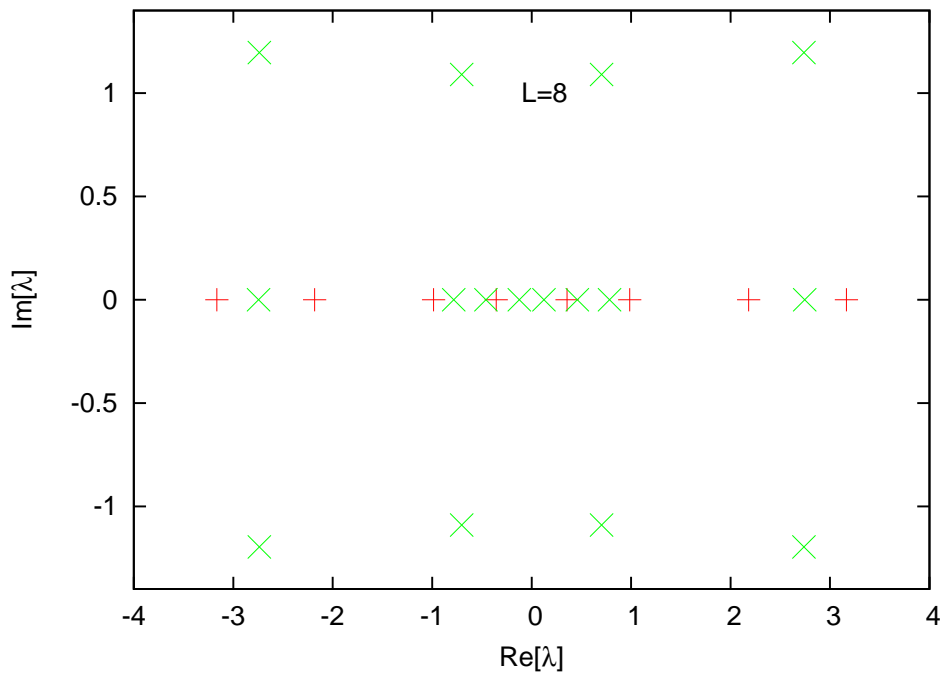}
\end{center}
\caption{The ground state Bethe roots $\lambda_j^{(1)}$ ({\color{green} $\times$}) and $\lambda_j^{(2)}$ ({\color{red} $+$}) for $L=8$.} 
\label{fig1}
\end{figure}

In order to get some insight on the spectrum of the $G_2$ spin chain we 
have diagonalized the Hamiltonian (\ref{HAM}) for lattice sizes $L \leq 8$.
The Hilbert space have been separated in many distinct sectors
exploring both the $U(1)$ charges
and the translation invariance due to periodic boundary conditions. For $L$ even,
we find that the ground state is a singlet occurring 
in the charge sector $(m_1,m_2)=(0,0)$. We next solve the Bethe equations (\ref{bethe}) in this
sector to obtain the structure of the corresponding ground state roots. In Fig.(\ref{fig1})
we exhibit our findings for the pattern of the ground state roots with size $L=8$.  

\begin{figure}[ht]
\begin{center}    
\includegraphics[width=13cm]{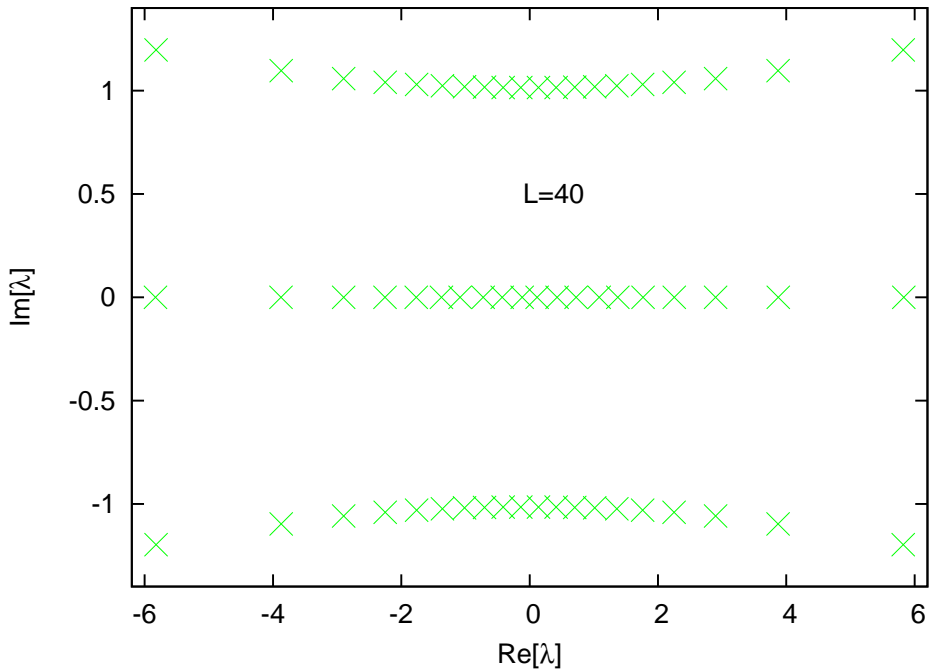}
\end{center}
\caption{The three-string branch of the root $\lambda_j^{(1)}$ for the size $L=40$.}
\label{fig2}
\end{figure}

We observe that the first level root $\lambda_j^{(1)}$ has two distinct branches, 
one of them lying solely on
the real axis (one-string type) while the other are made of complex roots having the typical 
form of a three-string 
$\{x_j \pm i y_j,x_j\}$. By way of contrast the second level roots $\lambda_j^{(2)}$ are always 
dominated by real roots. Assuming this pattern we have solved the Bethe equations for larger systems
to investigate the behavior of the complex part of the roots. In Fig.(\ref{fig2}) we show 
this pattern
for $L=40$ suggesting that as $L$ grows the imaginary part tends to 
the value $\pm i$. Therefore, we expect that in the thermodynamic limit 
the string hypothesis for
the ground state should be,
\begin{equation}
\lambda^{(1)}= \{ \xi_j^{(1)}, \xi_j^{(3)} \pm i, \xi_j^{(3)} \},~~~
\lambda^{(2)}= \{ \xi_j^{(2)} \}
\label{string}
\end{equation}
where the variables $\xi_j^{(1)}$, $\xi_j^{(2)}$ and $\xi_j^{(3)}$ are real numbers. At this point we recall
that this situation
is similar to that found for the $Sp(4)$ spin chain \cite{MAR2}. However, for the $Sp(4)$ spin chain we have a sea of two-string roots 
rather than a sea of three-string roots.

In order to study the thermodynamic limit we substitute 
the string assumption (\ref{string}) into the Bethe 
equations (\ref{bethe}) leading us
to three coupled non-linear equations for the real 
variables $\xi_j^{(l)}$.  
By taking the logarithm of the these equations we obtain,
\begin{eqnarray}
\label{bethestring}
L\phi_{1/2}\left(\xi_{j}^{(1)}\right)& =& 2 \pi \mathrm{Q}_{j}^{(1)}
+\sum_{\stackrel{k=1}{k \neq j}}^{M_{1}}
\phi_{1}\left(\xi_{j}^{(1)} - \xi_{k}^{(1)}\right) 
-\sum_{k=1}^{M_{2}}
\phi_{\frac{3}{2}}\left(\xi_{j}^{(1)} - \xi_{k}^{(2)}\right)  \nonumber \\
&+& \sum_{k=1}^{M_{3}} \left[
\phi_{1}\left(\xi_{j}^{(1)} - \xi_{k}^{(3)}\right) 
+\phi_{2}\left(\xi_{j}^{(1)} - \xi_{k}^{(3)}\right) \right],~~j=1,\dots,M_1, \nonumber \\
2\pi \mathrm{Q}_{j}^{(2)} +
\sum_{\stackrel{k=1}{k \neq j}}^{M_{2}}
\phi_{3}\left(\xi_{j}^{(2)} - \xi_{k}^{(2)}\right) &=&
\sum_{k=1}^{M_{3}} \left[
\phi_{\frac{5}{2}}\left(\xi_{j}^{(2)} - \xi_{k}^{(3)}\right) 
+\phi_{\frac{3}{2}}\left(\xi_{j}^{(2)} - \xi_{k}^{(3)}\right) 
+\phi_{\frac{1}{2}}\left(\xi_{j}^{(2)} - \xi_{k}^{(3)}\right) 
\right] \nonumber \\
&+&\sum_{k=1}^{M_1}\phi_{\frac{3}{2}}\left(\xi_{j}^{(2)} - \xi_{k}^{(1)}\right),~~j=1,\dots,M_2, \nonumber \\
L\phi_{\frac{3}{2}}\left(\xi_{j}^{(3)}\right)&=&2 \pi \mathrm{Q}_{j}^{(3)}
+\sum_{\stackrel{k=1}{k \neq j}}^{M_{3}} \left[
2\phi_{1}\left(\xi_{j}^{(3)} - \xi_{k}^{(3)}\right) 
+2\phi_{2}\left(\xi_{j}^{(3)} - \xi_{k}^{(3)}\right) 
+\phi_{3}\left(\xi_{j}^{(3)} - \xi_{k}^{(3)}\right) 
\right] \nonumber \\
&-& \sum_{k=1}^{M_{2}} \left[
\phi_{\frac{5}{2}}\left(\xi_{j}^{(3)} - \xi_{k}^{(2)}\right) 
+\phi_{\frac{3}{2}}\left(\xi_{j}^{(3)} - \xi_{k}^{(2)}\right) 
+\phi_{\frac{1}{2}}\left(\xi_{j}^{(3)} - \xi_{k}^{(2)}\right) 
\right] \nonumber \\
&+& \sum_{k=1}^{M_{1}} \left[
\phi_{1}\left(\xi_{j}^{(3)} - \xi_{k}^{(1)}\right) 
+\phi_{2}\left(\xi_{j}^{(3)} - \xi_{k}^{(1)}\right) \right],~~j=1,\dots,M_3 \nonumber \\
\label{nonlin}
\end{eqnarray}
where $\phi_{s}(x)=2\arctan(x/s)$ and the factors $2 \pi  \mathrm{Q}_{j}^{(l)}$ originates from the indefiniteness of
the logarithm branches. The numbers $Q_j^{(l)}$ are usually given in terms of sequences of integers or of half-integers 
and for the ground state their 
configuration are,
\begin{eqnarray}
\mathrm{Q}_j^{(1)} &=& (L/2-1)/2 -(j-1),~~j=1,\dots,L/2 \nonumber \\
\mathrm{Q}_j^{(2)} &=& (L-1)/2 -(j-1),~~j=1,\dots,L  \nonumber \\
\mathrm{Q}_j^{(3)} &=& (L/2-1)/2 -(j-1),~~j=1,\dots,L/2
\end{eqnarray}

We can take the thermodynamic limit following the method devised by 
Yang and Yang \cite{YANYAN}. In the $L \rightarrow \infty$ limit one expects that 
the Bethe roots fill the entire real axis with densities
$\rho^{(l)}(\xi)$ associated to the rapidities $\{\xi_j^{(l)}\}$ for $l=1,2,3$. Introducing the counting 
functions $z_j^{(l)}= Q_j^{(l)}/L$ such densities
are obtained by the derivatives,
\begin{equation}
\rho^{(l)}(\xi) = \frac{d z^{(l)}(\xi)}{d\xi},~~l=1,2,3
\end{equation}

By taking the derivatives of Eqs.(\ref{nonlin}) we find that 
such densities satisfy 
the following set of coupled linear integral
equations,    
\begin{equation}
2\pi \rho^{(l)}(\xi)+\sum_{j=1}^{3}
\int_{-\infty}^{+\infty} \mathrm{K}_{l,j}(\xi-\xi^{'}) \rho^{(j)}(\xi^{'}) d \xi^{'}=
\psi_{1/2}(\xi)\delta_{l,1}+ \psi_{\frac{3}{2}}(\xi)\delta_{l,3},~~~l=1,2,3
\end{equation}
where $\psi_s(\xi)=2s/(s^2+\xi^2)$ and the matrix elements of the kernel are,
\begin{equation}
K(\xi)=\left( \begin{array}{ccc} 
	\psi_1(\xi)  &  -\psi_{\frac{3}{2}}(\xi) & \psi_{1}(\xi) +\psi_{2}(\xi)  \\
	-\psi_{\frac{3}{2}}(\xi)  &  \psi_{3}(\xi) & -\psi_{\frac{5}{2}}(\xi)-\psi_{\frac{3}{2}}(\xi) -\psi_{\frac{1}{2}}(\xi)  \\
	\psi_{1}(\xi)+\psi_{2}(\xi)  &  -\psi_{\frac{5}{2}}(\xi)-\psi_{\frac{3}{2}}(\xi) -\psi_{\frac{1}{2}}(\xi) & 
	2\psi_{1}(\xi)+2\psi_{2}(\xi) +\psi_{3}(\xi)  \\
	\end{array}
	\right)
\end{equation}

These coupled integral equations can be 
solved by Fourier 
transformation and 
the final expressions for
the densities are,
\begin{eqnarray}
\label{dens}
&& \rho^{(1)}(\xi) = \frac{1}{2\cosh(\pi \xi)}, \nonumber \\
&& \rho^{(2)}(\xi) = \frac{1}{2\sqrt{3}} \frac{\cosh\left(\frac{\pi \xi}{6}\right)}{\cosh\left(\frac{\pi \xi}{3}\right) +\frac{1}{2}}, \nonumber \\
&& \rho^{(3)}(\xi) = \frac{1}{2\cosh(\pi \xi)} \left(1+ \sqrt{2} \sinh\left(\frac{\pi \xi}{2}\right) \sinh\left(\frac{\pi \xi}{3}\right)
-\sqrt{\frac{2}{3}} \cosh\left(\frac{\pi \xi}{2}\right) \cosh\left(\frac{\pi \xi}{3}\right) \right)
\end{eqnarray}

We now can compute the ground energy per site  
$e_{\infty}$ of the $G_2$ spin chain in the 
thermodynamic limit. 
By substituting the string hypothesis (\ref{string}) in Eq.(\ref{ener}) and after replacing the sums in the 
$L \rightarrow \infty$ limit by integrals we obtain,
\begin{eqnarray}
e_{\infty}= -\int_{-\infty}^{+\infty} \frac{\rho^{(1)}(\xi)}{\frac{1}{4}+\xi^2} d \xi 
-3\int_{-\infty}^{+\infty} \frac{\rho^{(3)}(\xi)}{\frac{9}{4}+\xi^2} d \xi +\frac{17}{12}
\end{eqnarray}
which can be  computed in terms of elementary functions and the final result is,
\begin{equation}
\label{exact}
e_{\infty}= -\frac{\pi (9-2\sqrt{3})}{18} -\frac{\log\left(2+\sqrt{3}\right)}{\sqrt{3}}+\frac{17}{12} \cong-0.309875868
\end{equation}

This result can be compared with the extrapolation of the ground state energy
for several values of $L$. Let us denote by $E_0(L)$ the ground state energy of 
the $G_2$ spin chain for a given size $L$.
In Table (\ref{tab1}) we show the 
finite-size sequence $E_0(L)/L$ up to $L=40$ together with the 
extrapolated value. We note that 
the extrapolated 
value is indeed in good
agreement with the exact value (\ref{exact}).
\begin{table}
\begin{center}
\begin{tabular}{|c|c|} \hline
$L$  & $E_0(L)/L$   \\ \hline \hline
8   &  -0.340238195878784    \\ \hline 
12  &    -0.323295274854638    \\ \hline 
16  &    -0.317406034364185   \\ \hline 
20  &   -0.314689188614338    \\ \hline 
24  &    -0.313216017511915    \\ \hline 
28  &   -0.312328705698502    \\ \hline 
32  &   -0.311753216558468    \\ \hline 
36  &   -0.311358859169196    \\ \hline 
40  &   -0.311076880008779    \\ \hline
$\mathrm{Extrapolated}$ & -0.3098757($\pm 1$)  \\ \hline
\end{tabular}
\end{center}
\caption{Finite-size sequence $E_0(L)/L$ for the ground state energy and its extrapolated value.}
\label{tab1}
\end{table}

\section{Finite-size effects}

We start by recalling that low-lying excitations over the ground 
state are described by producing modifications on the configuration of the Bethe roots. 
This is achieved for instance with the introduction of holes by using
alternative choices of the quantum numbers $Q_j^{(l)}$. Let us denote by $\epsilon^{(l)}(\xi)$ and $p^{(l)}(\xi)$
the energy and the momenta measured from the ground state associated to changes on the roots $\xi_j^{(l)}$. 
By using standard methods developed for solvable models \cite{DES,TAKA}
we find that the
low momenta excitations are gapless with a linear dispersion 
$\epsilon^{(l)}(\xi) \sim v_l p^{(l)}(\xi)$. The respective sound velocity $v_l$ 
is given by,
\begin{equation}
\label{velo}
v_l =\Bigg{|}\frac{1}{\rho^{(l)}(\xi)} \frac{d \rho^{(l)}(\xi)}{d \xi}\Bigg{|}_{\xi=\infty}
\end{equation}

For large values of the spectral parameter $\xi$ we note that the densities (\ref{dens}) 
have two distinct asymptotic behavior. The first density decay as $\exp(-\pi \xi)$ while the other two
densities behave asymptotically as $\exp(-\frac{\pi \xi }{6} )$. This means that we have two distinct sound
velocities and from expression (\ref{velo}) we find,
\begin{equation}
v_1=\pi,~~v_2=\frac{\pi}{6}
\end{equation}

The fact that we have two different sound velocities tell us that 
the continuum limit of the $G_2$ spin chain 
is not strictly conformally invariant. This is the typical behaviour
found for the so-called Tomonaga-Luttinger liquids in which charge and spin density 
waves are gapless traveling with different velocities \cite{HAL}. For instance, examples in the realm
of integrable models are the 
repulsive Hubbard model with generic filling \cite{WOY,FRA} and more recently 
the fundamental $Sp(4)$ invariant spin chain \cite{MAR2}. In this situation, the underlying 
continuum limit
is normally expected to be described in terms of the semi-direct product of 
two conformal field theories. The critical properties such as the conformal anomaly
can still be extract from finite-size corrections predicted for 
Lorentz invariant models \cite{BLO,AFF} 
after trivial modifications. In particular, the ground state energy $E_0(L)$ 
for large system sizes is expected to behave as,
\begin{equation}
E_0(L) -Le_{\infty} \cong -\frac{\pi}{6 L}\left(c_1 v_1 +c_2 v_2 \right)
\label{CFT}
\end{equation}
where $c_1$ and $c_2$ are the central charges of the
conformal field theories.

From Eq.(\ref{CFT}) we can estimate the combination 
of central charges from the
large $L$ limit of the following sequence,
\begin{equation}
C(L)= -\left(E_0(L) -Le_{\infty} \right) \frac{6 L}{\pi^2} \cong c_1+\frac{c_2}{6}
\label{FNZ}
\end{equation}
where we have normalized Eq.(\ref{CFT}) by the Fermi velocity $v_1=\pi$.

In Table (\ref{tab2}) we show the above 
finite-size sequences for sizes up to $L=56$ together 
with the extrapolated value. We note that result of the extrapolation
is very close to $\frac{7}{6}$. This number can be obtained from the right 
hand side of Eq.(\ref{FNZ}) by choosing
the central charges of the two conformal theories to be
$c_1=c_2=1$. In this interpretation we considered 
that the $G_2$ spin chain commutes with two $U(1)$ conserved 
charges and that such
symmetries should be present in the continuum limit. 
We next recall that the typical field theory with $U(1)$ symmetry is  
Gaussian model whose underlying 
central charge is exactly $c=1$ \cite{CONF}. 
\begin{table}
\begin{center}
\begin{tabular}{|c|c|} \hline
$L$  & $C(L)$   \\ \hline \hline
   8   &    1.18131501496519   \\ \hline             
   12  &    1.17475001765000   \\ \hline           
   16  &    1.17190587653062    \\ \hline          
   20  &   1.17044526692352      \\ \hline         
   24  &   1.16958673640317      \\ \hline         
   28  &    1.16903151737795     \\ \hline           
   32  &     1.16864633268190     \\ \hline           
   36  &      1.16836426879827    \\ \hline            
   40  &     1.16814858516650      \\ \hline          
   44  &    1.16797765541174       \\ \hline         
   48  &     1.16783802371419      \\ \hline          
   52  &    1.16772093061884        \\ \hline        
   56  &    1.16762045564018         \\ \hline       
$\mathrm{Extrapolated}$ & 1.1667($\pm 1$)  \\ \hline
\end{tabular}
\end{center}
\caption{Finite-size sequence for the effective central charge and its extrapolated value.}
\label{tab2}
\end{table}

\begin{figure}[ht]
\begin{center}    
\includegraphics[width=13cm]{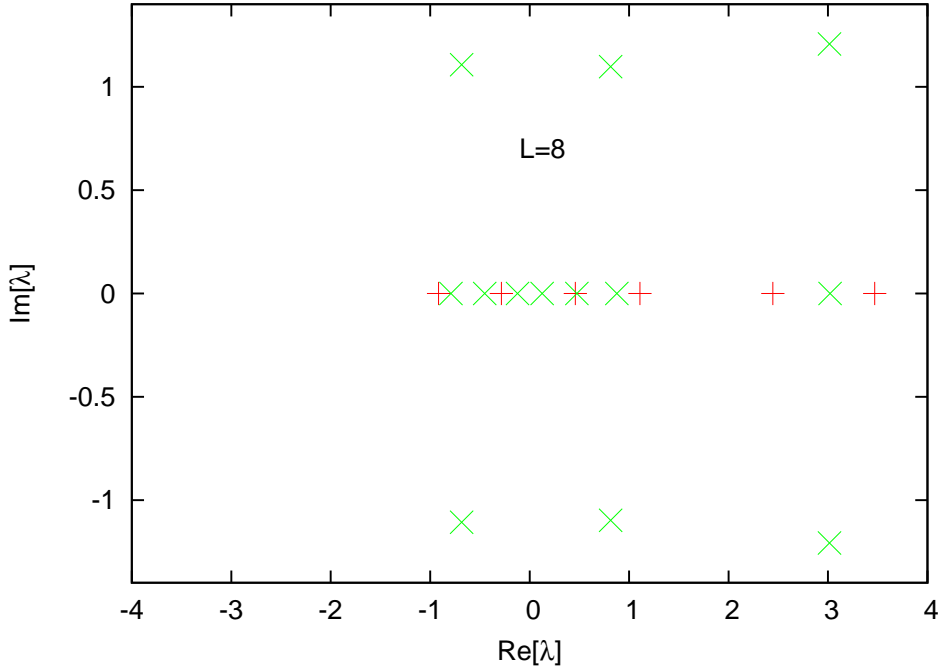}
\end{center}
\caption{The lowest excitation Bethe roots $\lambda_j^{(1)}$ ({\color{green} $\times$}) and $\lambda_j^{(2)}$ ({\color{red} $+$}) for $L=8$.} 
\label{fig3}
\end{figure}

We now turn our attention to the lowest excitation over the ground state. From our exact diagonalization 
we have identified that such state 
has non zero momenta being also twenty-eight fold degenerated. With finite Bethe roots we conclude that this state sits
in the sector $(m_1,m_2)=(1,1)$ and
the corresponding roots configuration are exhibited in Fig.(\ref{fig3}). We observe that this excitation is 
created by introducing one hole
on the sea of three-string roots and two holes on the second level real roots. By contrast, the number of the real 
roots on the first level remains the
same as that of the ground state. Considering our previous discussion on the behaviour of the dispersion relation we conclude 
that this type of excitation should be associated to density waves 
traveling with velocity 
$v_2=\frac{\pi}{6}$. 

We next investigate the finite-size correction to the energy $E_1(L)$ of the lowest excitation. 
The expected behaviour for the respective
energy gap measured from the ground state is,
\begin{equation}
E_1(L)-E_0(L) \cong \frac{2\pi}{L}(v_1 X_1 +v_2 X_2)
\label{FNZ1}
\end{equation}
where $X_1$ and $X_2$ are the anomalous dimensions of an operator defined on 
a product of two conformal field theories.

At this point we recall that this excitation does not involve modification on the pattern of the roots
associated to the density waves traveling with sound velocity $v_1=\frac{\pi}{2}$. It is therefore plausible to expect
that for the lowest excitation we should have $X_1=0$. In order to analyze the finite-size effects for
the lowest excitation we from Eq.(\ref{FNZ1})
built the following finite-size
sequences,
\begin{equation}
X(L)= \left(E_1(L) -E_0(L) \right) \frac{L}{2\pi^2} \cong X_1+\frac{X_2}{6}
\label{FINI}
\end{equation}

In Table (\ref{tab3}) we show the  
finite-size sequences (\ref{FINI}) for sizes up to $L=56$ together 
with the extrapolated value. We see that the extrapolation is close to one with rather 
good precision. This result strongly suggest that 
the lowest conformal dimension in operator content of the $G_2$ 
spin chain should be $X=X_2=1$. 
\begin{table}
\begin{center}
\begin{tabular}{|c|c|} \hline
$L$  & $X(L)$   \\ \hline \hline
   8   &    0.170089788970682   \\ \hline  
   12  &  0.169587643575080     \\ \hline
   16  &   0.169166699596788    \\ \hline 
   20  &    0.168868003281334    \\ \hline 
   24  &  0.168649566666569    \\ \hline 
   28  &     0.168483362545892   \\ \hline  
   32  &   0.168352529402626     \\ \hline
   36  &     0.168246649348275    \\ \hline 
   40  &   0.168159011240242    \\ \hline
   44  &     0.168085115914887    \\ \hline 
   48  &     0.168021839454732     \\ \hline
   52  &  0.167966946160385     \\ \hline
   56  &    0.167918793714462    \\ \hline
$\mathrm{Extrapolated}$ & 0.1667($\pm 1$)  \\ \hline
\end{tabular}
\end{center}
\caption{Finite-size sequence for the lowest conformal dimension and its extrapolated value.}
\label{tab3}
\end{table}

\section{The vertex model free-energy}

Let us denote by $\kappa(\lambda)$ the partition function 
per site of the vertex model on the square lattice $L \times L$
in the thermodynamic limit. From the fact that the partition 
function is the trace of $[\mathrm{T}_{L}(\lambda)]^{L}$ it follows,
\begin{equation}
\kappa(\lambda)= \lim_{L \rightarrow \infty} \left(\Lambda_{max}(L,\lambda)\right)^{1/L} 
\end{equation}
where $\Lambda_{max}(L,\lambda)$ denotes the largest eigenvalue 
of the transfer matrix $\mathrm{T}_{L}(\lambda)$.

The bulk free-energy per site in the thermodynamic limit $f_{\infty}(\lambda)$ 
is directly 
related to the partition function $\kappa(\lambda)$,
\begin{equation}
f_{\infty}(\lambda)=
\lim_{L \rightarrow \infty} \frac{1}{L} \log\left(\Lambda_{max}(L,\lambda)\right)= \log\left(\kappa(\lambda)\right)
\end{equation}

\begin{figure}[ht]
\begin{center}    
\includegraphics[width=13cm]{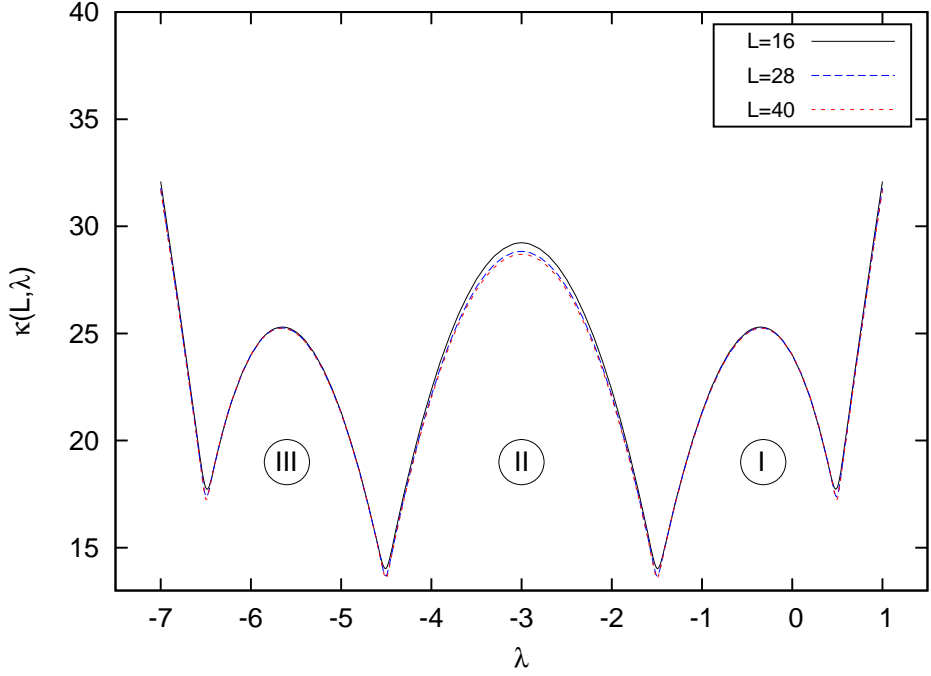}
\end{center}
\caption{The form of the partition function $\kappa(L,\lambda)$ for $L=16,28,40$.}
\label{fig4}
\end{figure}

Before considering the computation the bulk free-energy we shall first 
study the behaviour 
of the partition function per site for finite
values of $L$, namely
\begin{equation}
\kappa(L,\lambda)= \left(\Lambda_{max}(L,\lambda)\right)^{1/L} 
\end{equation}
which can be numerically computed by substituting the ground state Bethe roots
in the eigenvalue expressions (\ref{ein1}-\ref{ein3}). The behaviour of $\kappa(L,\lambda)$  
as a function of the spectral
parameter $\lambda$ is
illustrated in Fig.(\ref{fig4}) for $L=16,28,40$.

We observe that $\kappa(L,\lambda)$ is a limited function in the interval 
$-\frac{13}{2} -{\cal{O}}(e^{-L}) \leq \lambda \leq \frac{1}{2} +{\cal{O}}(e^{-L})$ 
and as the system size grows the partition function  develops sharp corners
around four points $\lambda= \frac{1}{2} \pm {\cal{O}}(e^{-L}), 
-\frac{3}{2} \pm {\cal{O}}(e^{-L}), 
-\frac{9}{2} \pm {\cal{O}}(e^{-L}),
-\frac{13}{2} \pm {\cal{O}}(e^{-L})$. We recall that this 
feature has recently been 
observed in the context
of the partition function of the $Sp(2N)$ 
vertex models in which was found two 
distinct spectral regimes \cite{JULI}.
In the case of the $G_2$ vertex model the limited region gives rise to three 
different regimes 
for the partition function  as indicated 
in Fig.(\ref{fig4}) by the symbols ${\bf I,II,III}$. In the thermodynamic 
limit we expect
that the spectral intervals of such regimes should be,
\begin{equation}
\label{regi}
{\bf{I}}:~~-\frac{3}{2} \leq \lambda \leq \frac{1}{2},~~~  
{\bf{II}}:~~-\frac{9}{2} \leq \lambda \leq -\frac{3}{2},~~~  
{\bf{III}}:~~-\frac{13}{2} \leq \lambda \leq -\frac{9}{2}  
\end{equation}

We next note that the Bethe roots associated to the largest transfer matrix 
eigenvalue is
symmetric around the origin. Therefore, from the property (\ref{cross})
the leading eigenvalue in the above regimes satisfy the following relations, 
\begin{eqnarray}
\Lambda_{max}^{({\bf{I}})}(L,\lambda) & =&  \Lambda_{max}^{({\bf{III}})}(L,-6-\lambda)~~\mathrm{for}~-\frac{3}{2} \leq \lambda \leq \frac{1}{2}, \nonumber \\
\Lambda_{max}^{({\bf{II})}}(L,\lambda) & =&  \Lambda_{max}^{({\bf{II}})}(L,-6-\lambda)~~\mathrm{for}~~-\frac{9}{2} \leq \lambda \leq -\frac{3}{2}
\label{prop}
\end{eqnarray}

We now turn our attention to the computation of the free-energy 
in the regimes (\ref{regi}) by using the results
established so far for the $G_2$ spin chain.

\subsection{Regimes ${\bf I}$ and ${\bf III}$} 

In the interval 
$-\frac{3}{2} \leq \lambda \leq \frac{1}{2}$ numerical calculations indicate that the 
leading eigenvalue of the transfer 
matrix is dominated by the first dressing polynomial $G_{1}(\lambda,\{\lambda_j^{(1)}\})$.
Therefore, for large $L$ we can write,
\begin{equation}
\label{fIaux}
f_{L}^{({\bf I})}(\lambda) = \log\left(|a_0(\lambda)|\right)+ \frac{1}{L}\sum_{j=1}^{2L} 
\log \left( 
\frac{\lambda_{j}^{(1)}+i\lambda-\frac{i}{2}}  
 {\lambda_{j}^{(1)}+i\lambda+\frac{i}{2}} \right)+{\cal O}(e^{-L})
\end{equation}

In order to consider the thermodynamic limit we first substitute the string 
hypothesis (\ref{string}) in Eq.(\ref{fIaux}). By taking the
limit $L \rightarrow \infty $ the expression 
for the bulk free-energy becomes,
\begin{equation}
f_{\infty}^{({\bf I})}(\lambda) = \log\left(|a_0(\lambda)|\right)+ 
\int_{-\infty}^{+\infty} 
\log \left( 
\frac{\xi+i\lambda-\frac{i}{2}}  
 {\xi+i\lambda+\frac{i}{2}} \right) \rho^{(1)}(\xi) d\xi+
\int_{-\infty}^{+\infty} 
\log \left( 
\frac{\xi+i\lambda-\frac{3i}{2}}  
 {\xi+i\lambda+\frac{3i}{2}} \right) \rho^{(3)}(\xi) d\xi
\end{equation}

The next step in our computation is to express 
the integrals in terms of their 
Fourier transformation. As a result we obtain,
\begin{equation}
f_{\infty}^{({\bf I})}(\lambda) = \log\left(|a_0(\lambda)|\right) -
\int_{0}^{+\infty} \frac{\sinh(x\lambda) \exp(-x/2)}{x \cosh(x/2)} dx
-\int_{0}^{+\infty} \frac{\sinh(x\lambda) \exp(-3x/2) \cosh(x) }{x \cosh(x/2) \cosh(3x)} dx
\label{intreI}
\end{equation}

These integrals can be expressed in terms of the logarithmic of a combination 
among elementary and gamma functions. The technical
details of the calculations are presented in Appendix A and in what follows we present 
only the main result. 
Expressing the bulk-free energy as a logarithmic 
of a partition function,
\begin{equation}
f_{\infty}^{({\bf I})}(\lambda)=\log\left(\kappa^{({\bf I})}(\lambda) \right) 
\label{freeIA}
\end{equation}
we find that the expression  for
$\kappa^{({\bf I})}(\lambda)$ is,
\begin{eqnarray}
\kappa^{({\bf I})}(\lambda) &=& 1728 \cot\left( \frac{\pi}{4}+\frac{\pi\lambda}{12}\right) \nonumber \\
&\times &\frac{
\Gamma\left(\frac{7}{12}-\frac{\lambda}{12}\right)
\Gamma\left(\frac{1}{6}-\frac{\lambda}{12}\right)
\Gamma\left(-\frac{\lambda}{12}\right)
\Gamma\left(\frac{2}{3}+\frac{\lambda}{12}\right)
\Gamma\left(\frac{13}{12}+\frac{\lambda}{12}\right)
\Gamma\left(\frac{3}{2}+\frac{\lambda}{12}\right)}
{\Gamma\left(\frac{7}{12}+\frac{\lambda}{12}\right)
\Gamma\left(\frac{1}{6}+\frac{\lambda}{12}\right)
\Gamma\left(\frac{\lambda}{12}\right)
\Gamma\left(-\frac{1}{3}-\frac{\lambda}{12}\right)
\Gamma\left(\frac{1}{12}-\frac{\lambda}{12}\right)
\Gamma\left(\frac{1}{2}-\frac{\lambda}{12}\right)} 
\label{freeI}
\end{eqnarray}

Finally, the expression for bulk free-energy in the 
the interval $-\frac{13}{2} \leq \lambda \leq -\frac{9}{2}$ can be obtained
exploring the property (\ref{prop}). Once again by writing the bulk free-energy
in this regime as,
\begin{equation}
f_{\infty}^{({\bf III})}(\lambda)=\log\left(\kappa^{({\bf III})}(\lambda) \right) 
\end{equation}
it follows that the corresponding partition function is,
\begin{eqnarray}
\kappa^{({\bf III})}(\lambda) & = & 1728 \tan\left(\frac{3\pi}{4}+\frac{\pi\lambda}{12}\right) \nonumber \\
& \times & \frac{
\Gamma\left(\frac{7}{12}-\frac{\lambda}{12}\right)
\Gamma\left(\frac{1}{6}-\frac{\lambda}{12}\right)
\Gamma\left(1-\frac{\lambda}{12}\right)
\Gamma\left(\frac{1}{2}+\frac{\lambda}{12}\right)
\Gamma\left(\frac{2}{3}+\frac{\lambda}{12}\right)
\Gamma\left(\frac{13}{12}+\frac{\lambda}{12}\right)}
{\Gamma\left(\frac{7}{12}+\frac{\lambda}{12}\right)
\Gamma\left(\frac{1}{6}+\frac{\lambda}{12}\right)
\Gamma\left(1+\frac{\lambda}{12}\right)
\Gamma\left(-\frac{1}{2}-\frac{\lambda}{12}\right)
\Gamma\left(-\frac{1}{3}-\frac{\lambda}{12}\right)
\Gamma\left(\frac{1}{12}-\frac{\lambda}{12}\right)} 
\label{freeIII}
\end{eqnarray}

\subsection{Regime ${\bf II}$} 

In this regime our numerical analysis suggests that the largest eigenvalue is mainly dominate by the third dressing
polynomial $G_{3}(\lambda,\{\lambda_j^{(1)},\lambda_j^{(2)}\})$. In what follows we shall assume that this term
will indeed dominate the thermodynamic limit in this regime. Following the procedure described above we obtain,
\begin{eqnarray}
f_{\infty}^{({\bf II})}(\lambda) & =& \log\left(|a_3(\lambda)|\right)+ 
\int_{-\infty}^{+\infty} 
\log \left( 
\frac{\xi+i(\lambda+\frac{5}{2})-i}  
 {\xi+i(\lambda+\frac{5}{2})+i} \right) \rho^{(1)}(\xi) d\xi \nonumber \\
&+& \int_{-\infty}^{+\infty} 
\log \left( 
\frac{\xi+i(\lambda+\frac{7}{2})+\frac{3i}{2}}  
 {\xi+i(\lambda+\frac{7}{2})-\frac{3i}{2}} \right) \rho^{(2)}(\xi) d\xi \nonumber \\
&+&
\int_{-\infty}^{+\infty} 
\log \left( 
\frac{(\xi+i(\lambda+\frac{5}{2})-i)  
(\xi+i(\lambda+\frac{5}{2})-2i)}  
{(\xi+i(\lambda+\frac{5}{2})+i) 
(\xi+i(\lambda+\frac{5}{2})+2i)} 
\right) \rho^{(3)}(\xi) d\xi
\end{eqnarray}
which can be further 
simplified by applying the Fourier transform method. It turns out that 
bulk free-energy can be rewritten as,
\begin{eqnarray}
f_{\infty}^{({\bf II})}(\lambda) & = &\log\left(|a_3(\lambda)|\right)- 
\int_{0}^{+\infty} \frac{\sinh((\lambda+\frac{5}{2})x) \exp(-x)}{x \cosh(x/2)} dx \nonumber \\
&+& 4\int_{0}^{+\infty} \frac{\cosh\left((\lambda+3)x\right) \sinh(x/2) \cosh(x) \exp(-3x/2)}{x \cosh(3x)} dx
\label{intereII}
\end{eqnarray}

The above integrals can be explicitly computed as explainned in Appendix A. As before by representing 
the bulk free-energy 
as a logarithmic function,
\begin{equation}
f_{\infty}^{({\bf II})}(\lambda)=\log\left(\kappa^{({\bf II})}(\lambda) \right) 
\end{equation}
we find that the corresponding partition function is given by,
\begin{eqnarray}
\kappa^{({\bf II})}(\lambda) & = & \frac{32}{\pi}\left(3^{(\frac{3}{2} -\frac{\lambda}{4})}\right ) \cot(\frac{\pi\lambda}{2}) 
\tan\left(\frac{3\pi}{4}+\frac{\pi\lambda}{12}\right) 
\tan\left(\frac{\pi}{6}+\frac{\pi\lambda}{12}\right) 
\nonumber \\
& \times & \frac{
\Gamma\left(\frac{7}{12}-\frac{\lambda}{12}\right)
\Gamma\left(\frac{1}{6}-\frac{\lambda}{12}\right)
\Gamma\left(\frac{5}{6}-\frac{\lambda}{12}\right)
\Gamma\left(1-\frac{\lambda}{12}\right)
\Gamma\left(-\frac{\lambda}{12}\right)
\Gamma\left(\frac{3}{2}+\frac{\lambda}{12}\right)
\Gamma\left(\frac{13}{12}+\frac{\lambda}{12}\right)}
{\Gamma\left(\frac{7}{12}+\frac{\lambda}{12}\right)
\Gamma\left(\frac{1}{12}-\frac{\lambda}{12}\right)
\Gamma\left(-1-\frac{\lambda}{4}\right)
\Gamma\left(\frac{1}{2}-\frac{\lambda}{12}\right)
\Gamma\left(1+\frac{\lambda}{12}\right)}
\label{freeII}
\end{eqnarray}
where one can verify that
$\kappa^{({\bf II})}(\lambda) = \kappa^{({\bf II})}(-6-\lambda)$ is indeed satisfied.

\begin{figure}[ht]
\begin{center}    
\includegraphics[width=13cm]{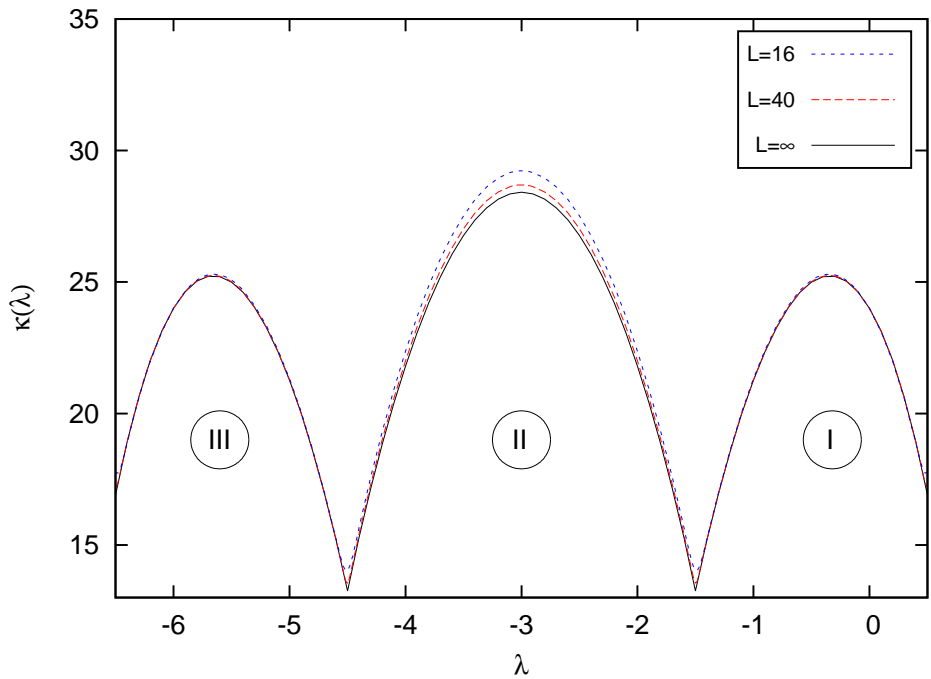}
\end{center}
\caption{Comparison of the partition function $\kappa(L,\lambda)$ for $L=16,40$ with that predicted for the thermodynamic limit.}
\label{fig5}
\end{figure}

For sake of comparison, in Fig.(\ref{fig5}) we show the numerical results  
for $\kappa(16,\lambda)$ and $\kappa(40,\lambda)$ together 
with the infinite volume values obtained by using
the expressions (\ref{freeI},\ref{freeIII},\ref{freeII})
in the interval $-\frac{13}{2} \leq \lambda \leq \frac{1}{2}$.
We observe that as the system size
grows the finite-size behaviour of partition function indeed go towards to that
predicted for the thermodynamic 
limit.

We finally note that the partition function is a continuous 
function on the spectral parameter
in the interval $-\frac{13}{2} \leq \lambda \leq \frac{1}{2}$.
However, the partition function is clearly not differentiable
at least at the points $\lambda=-\frac{3}{2},-\frac{9}{2}$. In fact, from 
Eqs.(\ref{freeI},\ref{freeIII},\ref{freeII}) one can check 
that the slopes 
at right and left sides of such points disagree. 

\section{Conclusions}

In this paper we have investigated several features associated to 
the spectrum of the $G_2$ invariant vertex model. We have formulated
the vertex model such that the $G_2$ symmetry becomes explicitly manifested. In particular, the respective $R$-matrix 
have been rewritten
in terms of the
quadratic Casimir of the $G_2$ Lie algebra. In addition, we have established a relationship among 
the numbers 
of Bethe roots entering the Bethe equations with the eigenvalues of the $U(1)$ conserved quantities 
associated to the Cartan subalgebra of $G_2$. 

We have identified the structure of the Bethe roots of the ground state of the $G_2$ spin chain. 
This makes possible the determination of the thermodynamic limit properties such as the ground state per site energy
of the spin chain. 
We have argued  that the low-lying excitations of the $G_2$ spin chain are massless but the respective density 
wave modes travel 
with distinct sound velocities. 
In order to provide some insights on the continuum limit of the spin chain we have studied the finite-size effects
to the ground state of the $G_2$ spin chain. From our numerical analysis we proposed that continuum limit should be
described in terms of the product of two $c=1$ conformal filed theories. We also have identified that the lowest excitation 
carries momenta and that the state is twenty-eight fold degenerated. We have analyzed the respective finite size corrections
and conjectured that the underlying conformal dimension should be $X=1$. A complete understanding of the operator content 
of the $G_2$ spin chain will require further investigation of the finite-size corrections 
for higher excited states. We hope to consider this problem in the context of the anisotropic $U_q[G_2]$ generalization 
of the spin chain in a future work. The presence of an extra parameter usually is able
to split degeneracies on the spectrum of the respective spin chain. This fact may help us to resolve
potential ambiguities on the interpretation of the conformal weights of the underlying field theory
which is expected to be not Lorentz invariant.

From the perspective of the $G_2$ vertex model we have indicated how the bulk free-energy can be computed by
a combination of numerical analysis and Bethe ansatz techniques. It turns out that 
we have been able to determine the bulk free-energy 
in a region of the spectral parameter in which the free-energy is limited function. Although the bulk free-energy
is continuous function on the spectral parameter we found that it has two sharp corner points in which 
differentiability is lost.

\section*{Acknowledgments}
This work was supported in part by the Brazilian Research Councils CNPq grant 305617/2021-4.

\addcontentsline{toc}{section}{Appendix A}
\section*{\bf Appendix A: Free-energy integrals}
\setcounter{equation}{0}
\renewcommand{\theequation}{A.\arabic{equation}}

The steps we have used to compute the integrals associated to 
the bulk free-energy 
are as follows. 
For the regime ${\bf I}$ we first use the following trigonometric identity,
\begin{eqnarray}
\frac{\cosh(x)}{\cosh(x/2) \cosh(3x)} &=& \frac{1}{\cosh(x/2)} +\frac{(2\sqrt{3}-3)}{12} \left( \frac{1}{\cosh(x/2)-\cos(\frac{\pi}{12})}+ 
\frac{1}{\cosh(x/2)-\cos(\frac{11\pi}{12})} \right) \nonumber \\
&-&
\frac{(2\sqrt{3}+3)}{12} \left( \frac{1}{\cosh(x/2)-\cos(\frac{5\pi}{12})}+ 
\frac{1}{\cosh(x/2)-\cos(\frac{7\pi}{12})} \right)
\label{AP1}
\end{eqnarray}

After using Eq.(\ref{AP1}) all the integrals entering the bulk free-energy are of the following form, 
\begin{equation}
\mathrm{I}(a,b,n,m)= \int_{0}^{+\infty} \frac{\exp(-ax) -\exp(-bx)}{x\left(\cosh(x)-\cos(\frac{m \pi}{n})\right)} dx
\end{equation}

It turns out that these integrals for $m$ odd can be expressed in terms of the logarithmic sum of gamma functions 
weighted by trigonometric factors,
\begin{equation}
\mathrm{I}(a,b,n,m)= \frac{2}{\sin\left(\frac{m\pi}{n}\right)} \sum_{k=1}^{n-1} \sin(\frac{m\pi k}{n})
\log\left(\frac{\Gamma(\frac{n+b+k}{2n}) \Gamma(\frac{n+a}{2n})}
{\Gamma(\frac{n+a+k}{2n}) \Gamma(\frac{n+b}{2n})} \right)
\label{int}
\end{equation}

By applying formula (\ref{int}) for each of the integrals of Eq.(\ref{intreI}) and 
after cumbersome simplifications
we obtain the final expressions (\ref{freeIA},\ref{freeI}). Recall that 
in the simplifications 
we have used 
standard properties of $\Gamma$-functions such as,
\begin{equation}
\tan(\pi x ) =\frac{\Gamma(\frac{1}{2}+x) \Gamma(\frac{1}{2}-x)}
{\Gamma(1-x) \Gamma(x)}
\end{equation}

The calculations for the regime ${\bf II}$ follows the same procedure described above. In this
case we have also used the following trigonometric identity,
\begin{equation}
\frac{\cosh(x)}{\cosh(3x)} = \frac{1}{4\sqrt{3}} \left( \frac{1}{\cosh(x)-\cos(\frac{\pi}{6})}- 
\frac{1}{\cosh(x)-\cos(\frac{5\pi}{6})} \right) \nonumber \\
\end{equation}

{}


\begin{thebibliography}{}
%
\bibitem{BAX} R.J. Baxter, {\em Exactly Solved Models in Statistical Mechanics}, Academic Press, New York, 1982.
%
\bibitem{YAN} C.N. Yang, {\em Phys.Rev.Lett. 19 (1967) 1312}; 
%
\bibitem{GAU} M. Gaudin, {\em Phys.Lett.A 24 (1967) 55}.
%
\bibitem{ZAMO} A.B. Zamolodichikov and Al.B. Zamolodchikov, {\em Ann.Phys. 120 (1979) 253}
%
\bibitem{KUL1} P.P. Kulish, N.Y. Reshetikin and E,K. Sklyanin, {\em Lett.Math.Phys. 5 (1981) 393}.
%
\bibitem{OGI} E.I. Ogievestky, {\em J.Phys.G: Nucl.Phys. 12 (1986) L105}.
%
\bibitem{FAD} L.A. Takhtajan and L.D. Faddeed, {\em Russ.Math.Surveys 34 (1979) 11}.
%
\bibitem{OGIV} E.I. Ogievesty and P. Wiegmann, {\em Phys.Lett.B 168 (1986) 360}.
%
\bibitem{RES1} N.Yu. Reshetikhin, {\em Theor.Math.Fiz. 63 (1985) 347}.
%
\bibitem{RES2} N.Yu. Reshetikhin, {\em Lett.Math.Phys. 14 (1987) 235}.
%
\bibitem{JUN} J. Suzuki, {\em Phys.Lett.A 195 (1994) 190}.
%
\bibitem{KOR} V.E. Korepin, G. Izergin and N.M. Bogoliubov, {\em Quantum Inverse Scattering Method, Correlation Functions and Algebraic
Bethe Anstaz}, Cambridge Univ. Press, Cambridge, 1992.
%
\bibitem{DEV} O. Babelon, H.J. de Vega and C.M. Viallet, {\em Nucl.Phys.B 200 (1982) 266} 
%
\bibitem{KUL} P.P. Kulish and N.Y. Reshetikhin, 
{\em J.Phys.A: Math.Gen. 16 (1983) L591}
%
\bibitem{MARA} M.J. Martins and P.B. Ramos, {\em Nucl.Phys.B 500 (1997) 579}.
%
\bibitem{VEGA} H.J. de Vega, {J.Phys.A:Math.Gen. 21 (1988) L1089} 
%
\bibitem{VESU} J. Suzuki, {\em J.Phys.A:Math.Gen. 21 (1988) L1175}
%
\bibitem{IZER} A.G. Izergin, V.E. Korepin and N.Yu. Reshetikhin, {\em J.Phys.A:Math.Gen. 22 (1989) 2615}
%
\bibitem{DELO} H.J. de Vega and E. Lopes, {\em Nucl.Phys. 362 (1992) 261}.
%
\bibitem{HUM}
J.E. Humphreys, {\em Lie algebras and representation theory}, Springer-Verlag, Heidelberg, 1972.
%
\bibitem{JAR} 
W. Fulton and J. Harris, {\em Representation Theory}, Springer-Verlag, Heidelberg, 1991.
%
\bibitem{OKU}
S. Okubo, {\em J.Math.Phys. 18 (1977) 2382}.
%
\bibitem{KUN} A. Kuniba, {\em J.Phys:Math.Gen. 23 (1990) 1349}.
%
\bibitem{BAT} C.M Yung and M.T. Batchelor, {\em Phys.Lett.A 198 (1995) 395}.
%
\bibitem{MAR2} M.J. Martins, {\em Nucl.Phys.B 636 (2002) 583}.
%
\bibitem{YANYAN} C.N.Yang and C.N. Yang, {\em J.Math.Phys. 10 (1969) 1115}.
%
\bibitem{DES} J. des Cloizeaux and J.J. Pearson, {\em Phys.Rev. 128 (1962) 2131}
%
\bibitem{TAKA} L.A. Takhtajan and L.D. Faddev, {\em Phys.Lett.A 85 (1981) 375}
%
\bibitem{HAL} F.D.M. Haldane, {\em J.Phys.C 14 (1981) 2589}, {\em Phys.Lett.A 81 (1981) 153}
%
\bibitem{WOY} F. Woynarovich, {\em J.Phys.A:Math.Gen.A 22 (1989) 4243}
%
\bibitem{FRA} H. Frahm and V.E. Korepin, {\em Phys.Rev.B 42 (1990) 10553}
%
\bibitem{BLO} H.W.J. Bl\"ote, J.L. Cardy and M.P. Nightingale, {\em Phys.Rev.Lett. 56 (1986) 742}
%
\bibitem{AFF} I. Affleck, {\em Phys.Rev.Lett. 56 (1986) 746}
%
\bibitem{CONF} P.Di. Francesco, P. Mathieu and D. Senechal, {\em Conformal Field Theory}, Springer-Verlag, Heidelberg, 1996
%
\bibitem{JULI} G.A.P. Ribeiro, A. K\"umper and P.A. Pearce, {\em J.Stat.Mech. (2022) 113102}, G.A.P. Ribeiro, {\em ArXiv: 2211.06487}
%
\end{thebibliography}
\end{document}